\documentclass[12pt]{article}
\usepackage[a4paper,margin=2cm]{geometry}
\usepackage[T1]{fontenc}
\usepackage{times}
\usepackage{enumerate}
\usepackage{amsmath,amssymb,amsthm,amsfonts}
\usepackage{graphicx}
\graphicspath{ {./images/} }
\usepackage{authblk}
\usepackage{color}
\usepackage[all,cmtip]{xy}
\usepackage[superscript]{cite} 
\usepackage{bm}
\usepackage{hyperref}
\usepackage{marginnote}
\usepackage{booktabs}
\usepackage{threeparttable}
\usepackage{bbm}
\usepackage{caption}
\usepackage{rotating}
\title{Do machine learning methods lead to similar individualized treatment rules? A comparison study on real data.}

\author[1]{Florie Bouvier \thanks{Correspondence to: Florie Bouvier (\href{mailto:florie.brion-bouvier@u-paris.fr}{florie.brion-bouvier@u-paris.fr})\\
Hôpital Hôtel-Dieu, 1 place du Parvis de Notre-Dame, 75004 Paris, France}}
\author[1]{Etienne Peyrot}
\author[1]{Alan Balendran}
\author[2]{Corentin S\'egalas}
\author[3]{Ian Roberts}
\author[1]{Fran\c{c}ois Petit}
\author[1,4]{Rapha\"el Porcher}

\affil[1]{Universit\'e Paris Cit\'e and Universit\'e Sorbonne Paris Nord, Inserm, INRAE, Center for Research in Epidemiology and StatisticS (CRESS), F-75004 Paris, France
}
\affil[2]{Universit\'e Bordeaux, Inserm, Bordeaux Population Health Research Center, Bordeaux, France}
\affil[3]{Clinical Trials Unit, London School of Hygiene \& Tropical Medicine, London, UK}
\affil[4]{Centre d'\'Epid\'emiologie Clinique, Assistance Publique-H\^opitaux de Paris, H\^otel-Dieu, Paris, France}

\date{}
\begin{document}
	
\theoremstyle{plain} 
\newtheorem*{theorem*}{Theorem} 
\newtheorem{theorem}{Theorem}[section]
\newtheorem*{corollary*}{Corollary} 
\newtheorem{corollary}[theorem]{Corollary}
\newtheorem{proposition}[theorem]{Proposition}
\newtheorem*{lemma*}{Lemma} 
\newtheorem{lemma}[theorem]{Lemma}
\theoremstyle{definition} 
\newtheorem{definition}[theorem]{Definition}
\newtheorem{example}[theorem]{Example}
\newtheorem{remark}[theorem]{Remark}
\newtheorem{assumption}{Assumption}
\newtheorem{examples}[theorem]{Examples}
\newtheorem{question}[theorem]{Question}
\newtheorem{Rem}[theorem]{Remark}
\newtheorem{Notation}[theorem]{Notations}

\counterwithout{equation}{section}

\maketitle

\begin{abstract}
Identifying patients who benefit from a treatment is a key aspect of personalized medicine, which allows the development of individualized treatment rules (ITRs). Many machine learning methods have been proposed to create such rules. However, to what extent the methods lead to similar ITRs, i.e., recommending the same treatment for the same individuals is unclear. In this work, we compared 22 of the most common approaches in two randomized control trials. Two classes of methods can be distinguished. The first class of methods relies on predicting individualized treatment effects from which an ITR is derived by recommending the treatment evaluated to the individuals with a predicted benefit. In the second class, methods directly estimate the ITR without estimating individualized treatment effects. For each trial, the performance of ITRs was assessed by various metrics, and the pairwise agreement between all ITRs was also calculated. 
Results showed that the ITRs obtained via the different methods generally had considerable disagreements regarding the patients to be treated. A better concordance was found among akin methods. Overall, when evaluating the performance of ITRs in a validation sample, all methods produced ITRs with limited performance, suggesting a high potential for optimism. For non-parametric methods, this optimism was likely due to overfitting. 
The different methods do not lead to similar ITRs and are therefore not interchangeable. The choice of the method strongly influences for which patients a certain treatment is recommended, drawing some concerns about their practical use.

\end{abstract}

\section{Introduction}\label{sec:intro}

Personalized medicine aims at tailoring a treatment strategy to the individual characteristics of each patient. An essential part of personalized medicine is identifying patients benefiting from a given treatment which allows the construction of individual treatment rules (ITRs). Briefly, ITRs are decision rules that recommend treatment based on patients' characteristics. Of particular interest are optimal treatment rules, which are rules that would lead to the best average outcome in the population if they were followed by all individuals \cite{Zhang2012}.\\
ITRs can be developed using data from randomized controlled trials (RCTs) or observational data. For instance, Farooq et al. developed the SYNTAX score II to guide decision-making between coronary artery bypass graft surgery (CABG) and percutaneous coronary intervention (PCI) in patients with complex coronary artery disease using data from the SYNTAX trial \cite{Farooq2013}. In this paper, we decided to focus on ITRs built from RCTs' data to avoid having to additionally handle confounding factors. However, all the approaches presented here could also be used with observational data.\\
The PATH statement outlines guidelines for conducting predictive analyses of heterogeneous treatment effects (HTE) in clinical trials. It establishes criteria for predicting HTE and thus developing ITRs, emphasizing the use of a risk modeling approach \cite{Kent_path,Kent2020}. However, alternative approaches have also been employed for predicting HTE \cite{rekkas_predictive_2020-1}, and a myriad of methods whose goal is to construct an ITR has been proposed in the last decade. Nonetheless, their relative performance is not clearly established and, more importantly, it is not clear whether the derived ITRs would lead to recommending the same treatment for the same individuals. This issue is worth studying because if the ITRs are not similar, it is important to know upstream when choosing a method to derive an ITR in real life.\\ 
From a conceptual viewpoint, two classes of methods to develop an ITR can be distinguished. The first class relies on deriving individualized treatment effects (ITE) and then an ITR by recommending treatment to those with a predicted benefit. This class can be further divided into two sub-classes: methods estimating the response surfaces and methods directly estimating the ITE via a contrast function. The second class comprehends methods that directly estimate the ITR without explicitly relying on estimating ITEs or a contrast function.\\
Some comparisons of methods constructing an ITR via the ITE have been performed in the past. In particular, Jacob and Zhang et al. have both studied the performance of meta-learners (T-learner, S-learner, X-learner, DR-learner and R-learner) and causal forests \cite{jacob_cate_2021,zhang_unified_2021}. Jacob found that the methods resulted in differences in terms of ITEs estimates and recommended using multiple methods and comparing their results in practice \cite{jacob_cate_2021}. In their paper, Zhang et al. also found that the methods performed differently \cite{zhang_unified_2021}. To our knowledge, no comparison has included all the methods we are presenting, particularly methods that directly construct an ITR without calculating ITEs. Furthermore, none of the previous works implemented metrics to assess the agreements between pairs of methods.\\
In this study, we aimed to compare a wide range of methods to develop ITRs, both in terms of performance and agreement. We compared 22 different methods and applied those using data from two randomized controlled trials: the International Stroke Trial (IST) and the CRASH-3 trial. The remainder of this paper is organized as follows. We start by introducing the statistical setting in section~\ref{sec:statset}. In section~\ref{sec:algos}, the different methods are presented. In section~\ref{sec:comp}, the 22 methods are compared on two real RCTs. Section~\ref{sec:discuss} concludes with a discussion.

\section{Statistical setting}\label{sec:statset}
In this section, the potential outcomes framework is introduced. Then, we explain how to construct an individualized treatment rule (ITR). Finally, we enumerate the metrics used to compare the ITRs.

\subsection{Causal framework}\label{ssec:causal}
We follow Rubin's potential outcomes framework \cite{Rubin1974}.
We assume access to an independent and identically distributed sample of observations. Let $X \in \mathcal{X}\subset\mathbb{R}^n$ a vector of covariate in the covariate space $\mathcal{X}$, $A \in \{0,1\}$ be an indicator variable for the treatment of interest and $Y \in \{0,1\}$ be a binary outcome. We introduce potential outcomes $Y^{0}$ and $Y^{1}$ that represent the binary outcomes that would be observed if patients were assigned to either the control or the evaluated treatment respectively. Without loss of generality, we assume that $Y=1$ is a desirable event. 
We make the following assumptions \cite{Goetghebeur2020}:
\begin{itemize}
    \item Consistency: the observed outcome corresponds to the potential outcome i.e. if a patient received the treatment their observed outcome would be $Y^{1}$ and if they received the control, their observed outcome would be $Y^{0}$.
    \item No interference: the outcome only depends on the treatment applied to the patient, and not on the treatment applied to other patients.
    \item Unconfoundedness: all characteristics associated with both the treatment assignment and the outcome, should have been measured in the study.
    \item Positivity: all patients have a non-null probability of receiving either treatment.
\end{itemize}
In the setting of RCTs, Unconfoundedness and Positivity are met by design.

\subsection{Individualized treatment rules}\label{ssec:itr}
We are interested in constructing individualized treatment rules (ITR) which are decision rules that recommend treatment based on patients' characteristics.\\
Those rules are modeled as maps $\begin{array}{ccccc}
r & : & \mathcal{X} & \to & \{0,1\}.
\end{array}$ Accordingly, for a given set of covariates $x \in \mathcal{X}$, $r(x)$ indicates whether or not the treatment should be given to a patient. An optimal rule $r^{opt}$ is obtained when the value $\mathcal{V}(r)$ among all $r \in \mathcal{R}$, with $\mathcal{R}$ being the class of all treatment rules, is maximized \cite{Zhang2012}:
$$r^{opt} = \underset{r \in \mathcal{R}}{\arg \max }\:\mathcal{V}(r),$$
where $\mathcal{V}(r) = E[Y(r)]$ with $Y(r) = Y^{1}r(x) + Y^{0}[1-r(x)]$ representing the outcome observed if the rule $r$ was followed.\\
Constructing an optimal treatment rule can be achieved in two different ways.
The first approach involves calculating individual treatment effects (ITE). The ITE $\tau$ represents the predicted benefit under one treatment minus the predicted benefit under the other treatment, given a set of patients’ characteristics:
$$\tau(x) = E(Y^{1} - Y^{0}|X = x) = \mu_1(x) - \mu_0(x).$$
An optimal rule is obtained by only giving the evaluated treatment to patients with a positive value of $\tau(x)$ i.e. $r^{opt}(x) = \mathbbm{1}_{\{\tau>0\}}(x)$. In this approach, some methods estimate the ITE by estimating the response surfaces whereas others directly estimate the ITE via a contrast function. 
The second approach consists of directly developing an optimal rule, without estimating the ITEs, by minimizing a loss function of the value of the rule. The methods to develop ITRs considered in this project are described in section~\ref{sec:algos}.

\subsection{Metrics}\label{ssec:metrics}
Several metrics were used to compare the ITRs developed with different methods described in section~\ref{sec:algos}. Using several metrics allows us to have a comprehensive view of the performance of the ITRs. Two classes of metrics can be distinguished: metrics whose aim is to estimate the performance of the rules and metrics whose aim is to compare the level of agreement between two rules.

\subsubsection{Performance metrics}\label{sssec:eval}
First, metrics to assess the quality of a single ITR were used, enabling us to compare the performance of the ITRs.
\begin{itemize}
    \item The value of a rule: As stated previously, the value $\mathcal{V}(r) = E[Y(r)]$ represents the mean outcome if the ITR was correctly followed. In this project, a desirable binary outcome is considered, thus, ITRs with $\mathcal{V}(r)$ closer to $1$ have a better performance.
    \item The benefit of the rule in terms of assigned treatment among people with a positive and negative score, is assessed with two metrics: $B_{pos}$ and $B_{neg}$, where $B_{pos}$ represents the average benefit of giving the evaluated treatment among people with a positive score i.e $r(x) = 1$ and $B_{neg}$ represents the average benefit of not giving the evaluated treatment among people with a negative score i.e $r(x) = 0$ \cite{janes_statistical_nodate}. The values are between $-1$ and $1$, with $1$ meaning there is a benefit in treating people with a positive score for $B_{pos}$ and a benefit in not treating people with a negative score for $B_{neg}$.
    $$B_{pos}=P(Y=1 \mid A=1, r(x) = 1)-P(Y=1 \mid A=0, r(x) = 1),$$
    $$B_{neg}=P(Y=1 \mid A=0, r(x) = 0)-P(Y=1 \mid A=1, r(x) = 0).$$
    \item The Population Average Prescription Effect (PAPE): PAPE compares an ITR with a treatment rule that randomly treats the same proportion of patients \cite{imai_experimental_2021}:
    $$PAPE= E[Y(r)-p_r Y^1-(1-p_r)Y^0]$$
    where $p_r$ represents the proportion of patients assigned to the evaluated treatment under the ITR $r$.\\
    The PAPE takes values between $-1$ and $1$. Here, since higher values of the outcome are desirable, higher values of PAPE indicate a better performance of the ITR. A value of $0$ indicates that the ITR does not perform better than treating randomly the same proportion of patients. Negative values mean that the ITR performs worse. An advantage of the PAPE is that it is easy to interpret.
    \item The c-statistic for benefit: it is the probability that from two randomly chosen matched pairs with unequal observed benefit, the pair with greater observed benefit also has a higher predicted probability where the observed benefit refers to the difference in outcomes between two patients with the same predicted benefit but with different treatment assignments \cite{vanKlaveren2018}. To create the pairs, a patient in the control group is matched to one in the treatment group with a similar predicted treatment benefit. Higher values of the c-statistic for benefit are better. The c-statistic for benefit quantifies how well the rule discriminates patients benefiting from patients not benefiting from taking a given treatment. The c-statistic for benefit can only be calculated for methods using an ITE or a benefit score to derive an ITR.
\end{itemize}
The standard errors for each metric were calculated through a Bootstrap procedure involving 1000 samples of the original dataset.

\subsubsection{Agreement between two rules}\label{sssec:comp}
Metrics to see if two ITRs have the same recommendation and agree to allocate the treatment to the same patients were used.
\begin{itemize}
    \item Matthews correlation coefficient (MCC): Here, the MCC is used to measure the disagreement, in terms of treated patients, between two rules \cite{baldi_assessing_2000}. The values range between $-1$ and $1$, where $1$ indicates a perfect positive correlation, $0$ indicates no correlation and $-1$ indicates a perfect negative correlation.
    \item Cohen's kappa coefficient: Cohen's kappa measures the agreement between two rules by considering the number of agreements and disagreements \cite{cohen_coefficient_1960}. It can range from $-1$ to $1$. A value inferior to $0$ demonstrates that there is less than chance agreement between the two rules, a value of $0$ shows no agreement and a value of $1$ means that there is perfect agreement between the rules.
\end{itemize}

\subsection{Multiple Correspondence Analysis}
A Multiple Correspondence Analysis (MCA) was conducted to see if the ITRs agreed on the treatment decision in the presence of some specific characteristics. All variables included in the different models were put in the MCA, as well as the treatment allocation recommended by each ITR. Continuous variables were categorized and the choice of the categories was motivated by previous works done using the datasets \cite{the_international_stroke_trial_collaborative_group_international_2011,crash3_effects_2019}.

\section{Methods to construct individualized treatment rules}\label{sec:algos}
\begin{figure}[h]
\centerline{\includegraphics[scale=1.7]{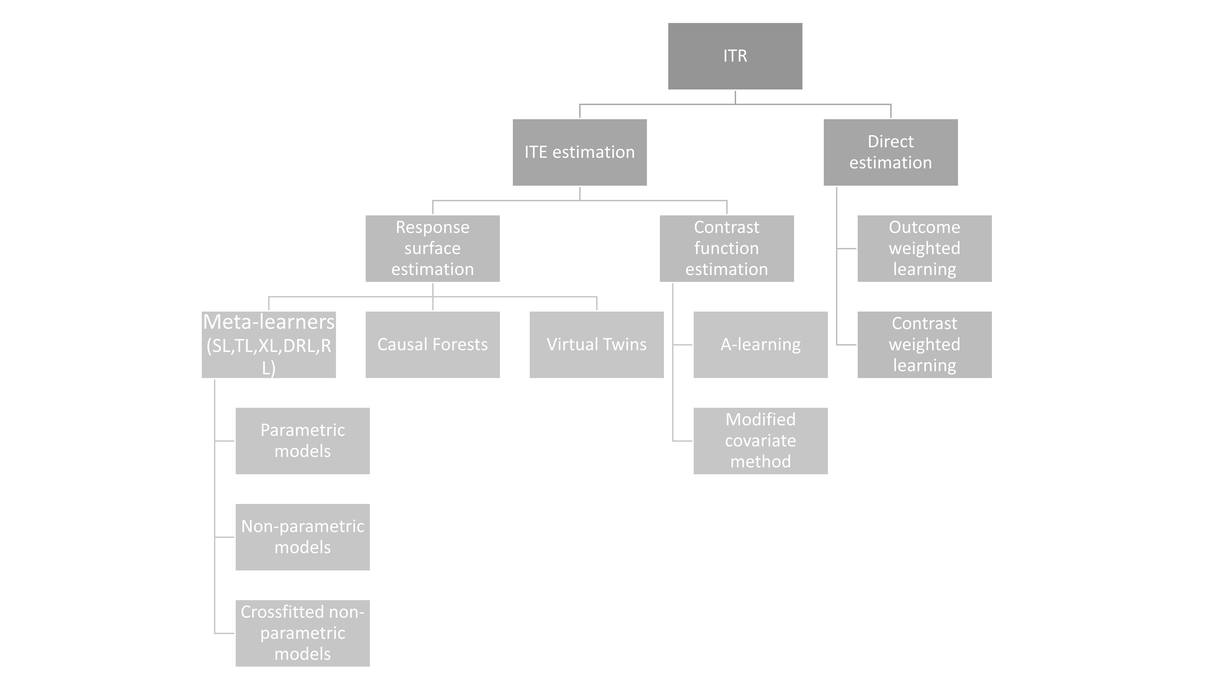}}
\caption{Classification of the methods.\label{fig:classif}}
\end{figure}

This section presents the different methods that were compared. We selected the most common methods which are either simple to implement with the R software \cite{R_software} or for which a package is available. As mentioned in Section~\ref{ssec:itr}, two classes of methods to develop an ITR were distinguished: a first class in which the ITR is constructed by first modeling the ITE, and an optimal rule is found by giving the treatment evaluated to the individuals with a positive ITE, and a second class, in which the ITR is directly estimated without calculating individualized treatment effects, and where an optimal rule is found by minimizing the risk of the value of the rule via a loss function. In the first class, two distinct approaches can be used to obtain the ITE: either estimating the expected difference of the potential outcomes between treatments or estimating the ITE directly via a contrast function. The majority of the methods fell under the first category: the meta-learners (T-learner, S-learner, X-learner, DR-learner, and R-learner, both with parametric and non-parametric models), PATH, causal forests, virtual twins, A-learning and the modified covariate method, whereas outcome weighted learning and contrast weighted learning fell under the second class. A classification of methods based on how they construct an optimal treatment rule is given in Figure~\ref{fig:classif}. Conceptually, some methods are related and are therefore referred to as belonging to the same family (e.g. parametric meta-learners, non-parametric meta-learners, A-learning, and the modified covariate method).

\subsection{Meta-learners}\label{ssec:meta}
Meta-learners are methods that use sub-regression problems to estimate the ITE via a base learner. In this project, two base learners were implemented: logistic regression and random forest (RF), the latter is selected for its good performance on tabular data \cite{grinsztajn_2022}. Since meta-learners can use several base learners, they are flexible and can adapt to different types of data.\\

\noindent Meta-learners with a non-parametric model as the base learner can be prone to overfitting. A solution to this potential overfitting is to use cross-fit \cite{chernozhukov_doubledebiasedneyman_2017}. Cross-fit consists of splitting the dataset into several folds. Then, the ITEs are learned on every fold and the results are aggregated to derive an ITR (more details in Supplementary Material~S1).
The meta-learners were compared with and without cross-fit when using random forests as a base learner.
When cross-fit was applied, 5 folds and 30 splits were used, because such a choice has been reported as leading to a good performance \cite{jacob_cross-fitting_2020}. 
However, to our knowledge, there is no standard method or clear guidance on how to perform cross-fit, and other choices exist \cite{jacob_cross-fitting_2020}.\\
In the upcoming segments, we use $\hat{\tau}$ to represent the estimate of $\tau$, and adhere to the same convention for denoting the estimates of other parameters.

\subsubsection{S-learner}\label{sl}
The S-learner estimates the treatment effect within a single regression model, where the treatment is included as a feature and where interactions between the treatment and relevant covariates are introduced in the parametric setting \cite{Kunzel2019}.
First, use a model to estimate the response function $\mu(x,a)$: $$\mu(x,a) = \mathbb{E}(Y|X=x,A=a).$$
Then, estimate the individual treatment effect $\tau$:
$$\widehat{\tau}(x) = \widehat{\mu}(x,1) - \widehat{\mu}(x,0).$$

\subsubsection{T-learner}\label{tl}
In the T-learner algorithm, two models are built, one in the treatment group and one in the control group\cite{Kunzel2019}. These models are used to calculate the response functions:
$$\mu_0(x) = \mathbb{E}(Y|X=x,A=0),$$
$$\mu_1(x) = \mathbb{E}(Y|X=x,A=1).$$
The ITE is estimated as the difference between the two predicted risks:
$$\hat{\tau}(x) = \widehat{\mu_1}(x) - \widehat{\mu_0}(x).$$

\subsubsection{X-learner}\label{xl}
The X-learner consists of three steps\cite{Kunzel2019}:
\begin{enumerate}
    \item Estimate the response functions as in the T-learner:
    $$\mu_0(x) = \mathbb{E}(Y|X=x,A=0),$$
    $$\mu_1(x) = \mathbb{E}(Y|X=x,A=1).$$
    \item Impute the treatment effects for the individuals in the treated group based on the control-outcome estimator and the treatment effects for the individuals in the control group based on the treatment-outcome estimator and estimate $\widehat{\tau_{1}}(x)$ and $\widehat{\tau_{0}}(x)$:
    $$\widetilde{D^{1}} = Y^{1} - \widehat{\mu_0}(X^{1}),$$
    $$\widetilde{D^{0}} = \widehat{\mu_1}(X^{0}) - Y^{0},$$
    $$\widehat{\tau_{1}}(x) = \mathbb{E}(\widetilde{D^{1}}|X = x),$$
    $$\widehat{\tau_{0}}(x) = \mathbb{E}(\widetilde{D^{0}}|X = x).$$
    \item Define the ITE by a weighted average of the two estimates:
    $$\widehat{\tau}(x) = w(x)\widehat{\tau_{0}}(x) + (1-w(x))\widehat{\tau_{1}}(x)$$
    where $w(x) \in [0,1]$ is a weighting function. An estimate of the propensity scores can be chosen as the weighting function, but there is no clear theory on how to choose an optimal weighting function. In the setting of RCTs, it is natural to choose $w(x) = \frac{1}{2}$ if the trial had a 1:1 randomization ratio.
\end{enumerate}
The X-learner has been described as advantageous in an unbalanced design or with sparse treatment effects \cite{Kunzel2019}.

\subsubsection{DR-learner}\label{drl}
The DR-learner is a doubly robust estimator that estimates the ITE in two stages \cite{Kennedy2020}. This learner includes double sample splitting to reduce bias.\\
First, the data $Z_i=(X_i,A_i,Y_i)$ are randomly split into three independent samples $D_{1a},D_{1b},D_2$. Then, the following two steps are applied:
\begin{enumerate}
    \item Construct propensity score estimates $\hat{\pi}$ of the propensity scores $\pi(X)= \mathbb{P}(A=1|X=x)$ using $D_{1a}$ and estimate the response functions $\widehat{\mu_0}$ and $\widehat{\mu_1}$ using $D_{1b}$.
    \item Construct the pseudo-outcome:
    $$\widehat{\varphi}(Z)=\frac{A-\widehat{\pi}(X)}{\widehat{\pi}(X)[1-\widehat{\pi}(X)]}[Y-\widehat{\mu}_A(X)]+\widehat{\mu}_1(X)-\widehat{\mu}_0(X).$$
    Then, regressing it on covariates $X$ of $D_2$ to estimate the ITE:
    $$\widehat{\tau}(x)=\widehat{\mathbb{E}}[\widehat{\varphi}(Z) \mid X=x].$$
\end{enumerate}
Cross-fitting can be added as an additional third step:
\begin{enumerate}\addtocounter{enumi}{2}
    \item Repeat steps 1 and 2 twice. First, $D_{1b}$ and $D_2$ are used for step 1 and $D_{1a}$ is used for step 2. Then, $D_{1a}$ and $D_2$ are used for step 1 and $D_{1b}$ is used for step 2. A final estimate of $\tau$ is constructed by averaging the three estimates.
\end{enumerate}
In this work, the propensity score was taken equal to $\frac{1}{2}$ since data from RCTs with a 1:1 randomization ratio were used.

\subsubsection{R-learner}\label{rl}
The R-learner estimates the ITEs in two steps \cite{Nie2020}:
\begin{enumerate}
    \item Fit the response function $\hat{\mu}(x)$ and the propensity scores $\hat{\pi}(x)$ with a base learner.
    \item Estimate ITEs by minimizing the R-loss, which uses Robinson's decomposition \cite{Robinson1988}:
    $$ 
    L_R(\tau(x))=\frac{1}{n} \sum_{i=1}^n[(Y_i-\hat{\mu}(X_i))-(A_i-\hat{\pi}(X_i)) \tau(X_i)]^2 + \Lambda_n(\tau(\cdot))
    $$
    where $\Lambda_n(\tau(\cdot))$ is a regularization term on the complexity of $\tau(\cdot)$.
\end{enumerate}
The response function and the propensity scores can be fitted using a cross-validation procedure and the regularization could be done with a penalized regression such as lasso or ridge, for instance, when logistic regression is used as a base learner \cite{Nie2020}. When random forests are used, regularization is achieved by tuning the hyperparameters, particularly by limiting the tree's depth and the number of variables used to build the trees.

\subsection{PATH approach}\label{ssec:path}
PATH is a risk modeling approach that has been recommended by the PATH statement \cite{Kent2020,Kent_path}. This method involves three steps:
\begin{enumerate}
    \item Fit a regression model (a logistic model with a binary outcome) with the relevant variables to derive the linear predictor
    \item Build a model that incorporates the linear predictor, the treatment variable, and the interaction between the linear predictor and treatment to estimate the response functions
    \item Derive the ITE based on the response functions
\end{enumerate}
This approach has demonstrated strong performance in diverse scenarios according to previous simulation studies \cite{rekkas_estimating_2023,van_klaveren_models_2019-2}.

\subsection{Causal forests}\label{ssec:cf}
The causal forests algorithm is a special case of generalized random forests (GRF), a flexible and general framework to estimate the ITEs \cite{athey_grf_2018}. 
Causal forests extend the original random forest algorithm by borrowing ideas from kernel-based methods and the R-learner \cite{Nie2020}.\\
In contrast to the standard random forest algorithm in which a prediction for a new observation is obtained by averaging predictions of each tree, here, the trees are used to compute a weighting scheme similar to kernel-based methods.
The trees act as weights between training points and any new observations:
$$\alpha_{bi}(x)=\frac{\mathbbm{1}_{X_i \in L_b(x)}}{|L_b(x)|}, \quad \alpha_i(x)=\frac{1}{B} \sum_{b=1}^B \alpha_{bi}(x)$$
where $X_i$ corresponds to the covariates of individual $i$ in the training dataset and $L_b(x)$ corresponds to the set of observations in the training set that fall in the same leaf as $x$ for tree $b$.\\
Then, the prediction for a new observation is obtained using the adaptive weights by minimizing the R-loss described above.\\
Another characteristic of causal forests (and more generally of GRF) is the notion of honesty where the training data is split into two parts: one for constructing the tree and the other (the estimation sample) for estimating leaf values for each tree. In doing so, the estimates are less prone to bias and more consistent. The notion of honesty is similar to employing the crossfit in non-parametric meta-learners.

\subsection{Virtual twins}\label{ssec:vt}
The virtual twins method consists in predicting response probabilities for treatment and control twins for all individuals using counterfactual models \cite{foster2011subgroup}. The difference in the probabilities is then used as the outcome in a classification or regression tree. A subgroup of individuals defined by a region $S$ of the covariate space $\mathcal{X}$ for which the treatment effect $\tau$ is better than a prespecified threshold can be then identified. The two steps are described below:
\begin{enumerate}
    \item Fit a random forest in which the covariates, the treatment indicator, and treatment-covariates interactions are included to estimate the response function $\mu$ and the ITE $\tau$ as in S-learner.
    \item Build a regression or a classification tree to find the covariates $X$ that are strongly associated with $\tau$ to define region $S$. Define $\tau*$, a binary variable, as the outcome. When $\tau>c$, $\tau*=1$ and when $\tau \leq c$, $\tau*=0$. Develop an ITR based on the value of $\tau*$. Individuals for which $\tau*=1$ are placed in the estimated region $\hat{S}$. The evaluated treatment is given to individuals in $\hat{S}$
\end{enumerate}
In this work, we built a classification tree and set $c$ equal to $0$.

\noindent The enhanced treatment effect Q(S) defined as:
$$Q(S) = (P(Y=1|A=1,X \in S)-P(Y=1|A=0,X \in S)) - (P(Y=1|A=1)-P(Y=1|A=0))$$ can be estimated by estimating $P(Y=1|A=1,X \in \hat{S})$, $P(Y=1|A=0,X \in \hat{S})$,  $P(Y=1|A=1)$ and $P(Y=1|A=0)$ using the observed proportions of the data. One of the different approaches that can be used to correct bias is bootstrapping. Bootstrapping measures the bias of $Q(\hat{S})$, which is then used to adjust $Q(\hat{S})$. In their work, Foster et al. compared bootstrapping to other approaches; their conclusion favored bootstrap with 20 samples  \cite{foster2011subgroup}.

\subsection{A-learning and the modified covariate method}\label{ssec:wal}
A-learning and the modified covariate method are two methods that focus on treatment-covariates interactions since treatment selection solely depends on the sign of the interactions \cite{Chen2017, tian2014predicting}.
Given the covariates and the treatment, the estimated outcome can be written as:
$$E(Y \mid A, X)=m(X)+ A\Delta(X)$$ 
where $m(X), \Delta(X)$ represent respectively the main effect of X and the treatment effect given X. Only the signs of $\Delta(X)$ matter for treatment selection.\\
In both methods, a personalized benefit score model $f$ is calculated and its sign, which is consistent with the direction of the treatment effect, is used to construct an ITR. An optimal ITR is found for both methods by minimizing a certain loss function $\ell$. Details on the loss functions are given below.

\subsubsection{A-learning}
In A-learning, the following expected loss function is considered: 
$$
\ell_{A}(f)=E(\ell_A(f, x))
$$ with
$$
\begin{aligned}
\ell_A(f, x)= & \pi(x) E[M(Y,(1-\pi(x)) f(x)) \mid A=1, X=x] \\
& + (1-\pi(x)) E[M(Y,-\pi(x) f(x)) \mid A=0, X=x] .
\end{aligned}
$$ where $\pi(x)$ represents the propensity scores and $M$ is a positive function, such as the quadratic or cross-entropy (also called logistic loss).\\
$\ell_A(f, x)$ is then replaced by its empirical version on the observed data:
$$
L_{A}(f)=\frac{1}{n} \sum_{i=1}^{n} M(Y_{i},(A_i-\pi(X_i)) f(X_i))
$$
$\pi(X_i)$ equals $\frac{1}{2}$ in the context of RCTs with 1:1 randomization.\\
When M is chosen to be the logistic loss, $L_{A}(f)$ is expressed as:
$$
L_{A}(f)=-\frac{1}{n} \sum_{i} Y_{i}(A_i-\pi(X_i)) f(X_i) -\log(1+\exp((A_i-\pi(X_i))  f(X_i))).
$$

\subsubsection{Modified covariate method}
Similarly, the expected loss function $\ell_{MCM}(f)=E(\ell_{MCM}(f, x))$  of the modified covariate method where
$$
\begin{aligned}
\ell_{MCM}(f,x) = & E[M(Y, f(X)) \mid A=1, X=x] \\
& +E[M(Y,-f(X)) \mid A=0, X=x]
\end{aligned}
$$
Its empirical version is
$$
L_{MCM}(f)=\frac{1}{n} \sum_{i=1}^{n} \frac{M(Y_i, (2A_i-1)f(X_i))}{(2A_i-1)\pi(X_i)+1-A_i},
$$
which boils down to
$$
L_{MCM}(f)=-\frac{2}{n} \sum_{i} Y_i(2A_i-1)f(X_i)-\log(1+\exp((2A_i-1)f(X_i))
$$
when the logistic loss function is used and $\pi(X_i)=\frac{1}{2}$. It is worth mentioning that by substituting the benefit score of A-learning in $L_A$ with double the benefit score of the modified covariate method, we obtain $L_{MCM} = 2L_A$.

\subsection{Outcome weighted learning}\label{ssec:owl}
Outcome weighted learning (OWL) uses a weighted classification framework, in which each patient is weighted based on their outcome, with a hinge loss to estimate an ITR \cite{Zhao2012}. An optimal treatment rule is obtained by minimizing the following quantity:
$$
L(f) = \frac{1}{n} \sum_{i=1}^n \frac{Y_i}{(2A_{i}-1) \pi(X_i)+1-A_i}(1-(2A_{i}-1) f(X_i))^{+}+\lambda_n\|f\|^2
$$
where $x^{+}=\max (x, 0)$, $\pi(X_i)$ represents the propensity scores, $\lambda_n$ is a penalty parameter used to avoid overfitting and $\|f\|$ is a norm for the function $\begin{array}{ccccc}
f & : & \mathcal{X} & \to & \{0,1\} \end{array}$. When employing the linear kernel, the Euclidean norm of the coefficients, excluding the intercept, is utilized. The parameter $\lambda_n$ is chosen by performing a cross-validation.
OWL is a consistent estimator and has low variability \cite{Zhao2012}.

\subsection{Contrast weighted learning}\label{ssec:cwl}
The idea behind contrast weighted learning (CWL) is to use contrasts of the outcome between pairs of patients to build weights used in a weighted classification algorithm to estimate an ITR \cite{guo_contrast_2022}. A contrast function $h$ is defined for a pair of patients to measure the relative favorability of their outcomes. Several contrast functions exist such as the difference $h(Y_i, Y_j)=Y_i-Y_j$, the log ratio $h(Y_i, Y_j)=\log(Y_i / Y_j)$ or the win indicator $h(Y_i, Y_j)=\operatorname{sgn}(Y_i-Y_j)$, with $\operatorname{sgn}(Y_i-Y_j)=1$ if $Y_i-Y_j>0$; $\operatorname{sgn}(Y_i-Y_j)=0$ if $Y_i-Y_j=0$ and $\operatorname{sgn}(Y_i-Y_j)=-1$ if $Y_i-Y_j<0$. In this project, we used the win indicator, considered the most robust contrast function by Guo et al. \cite{guo_contrast_2022}.
The optimal ITR is found by minimizing the following function:

\begin{equation*}
    L(f)=\frac{1}{2} E[(\mathbbm{1}_{h(Y_i, Y_j) (2A_i-1) f(X_i)<0}
    + \mathbbm{1}_{h(Y_i, Y_j) (2A_j-1) f(X_j) \geq 0}) \frac{|h(Y_i, Y_j)|}{\pi(X_i) \pi(X_j)}]
\end{equation*}

\noindent where $h(Y_i, Y_j)$ is the contrast function between patient $i$ and patient $j$ and $\pi(X_j)$ represents the propensity score. Here $h(Y_i, Y_j)=\operatorname{sgn}(Y_i-Y_j)$ and $\pi(X_i)=\pi(X_j)=1/2.$
CWL is a flexible and robust method that only relies on the contrast of outcomes between two patients. However, a correctly specified model is needed to ensure consistency.

\subsection{Implementation}
All the analyses were performed in R version 4.1.2. Virtual Twins was implemented using the \texttt{aVirtualTwins} package \cite{virtual_twins}. The package \texttt{personalized} was used to develop the modified covariate method and A-learning \cite{personalized}. For outcome weighted learning and contrast weighted learning, the package \texttt{WeightSVM} was used \cite{WeightSVM}. Causal forests was implemented using the package \texttt{grf} \cite{grf}. More details about the implementation such as the choice of the hyper-parameters can be found in Supplementary Material~S2.

\section{Comparison of the methods on real data}\label{sec:comp}
In this section, the ITRs obtained when applying the methods described above are compared on two multi-center randomized control trials: the International Stroke Trial and the CRASH-3 trial. The train and test datasets were obtained by splitting the data at the center level using $2/3$ of the data for training.

\subsection{International Stroke Trial}\label{ssec:ist}
The 22 methods were first compared on the International Stroke Trial (IST) \cite{the_international_stroke_trial_collaborative_group_international_2011}.The IST was chosen because an ITR has been developed on this dataset in the past using the T-learner method and found that 74\% of patients would benefit from taking aspirin \cite{Nguyen2020}. The IST is a multi-center randomized control trial that includes $19,435$ patients recruited from $466$ centers and examines the impact of administrating aspirin, heparin, or both in stroke. For our illustration, we focused on the impact of aspirin on stroke.
Nineteen variables were included in the different methods: 16 categorical variables and 3 continuous variables, similar to what Nguyen et al. did \cite{Nguyen2020}. The outcome used was death or dependency at 6 months (1=no and 0=yes). The treatment variable was binary (0=no aspirin and 1=aspirin). A description of the covariates and the outcome is reported in Supporting Material~S3.\\

\noindent Results of the metrics used to evaluate the ITR produced by each method are given in Table~\ref{tab:res_ist}. The performance in the train dataset of IST can be found in Supplementary Material~S3.
Higher values of c-statistic for benefit are better but it is rare to obtain values above $0.6$ \cite{vanKlaveren2018}. Here, the c-statistic for benefit was close to $0.5$ for all the ITRs, indicating poor discrimination. A reason for the poor discrimination could be the lack of strong heterogeneous treatment effects. This hypothesis was confirmed by conducting a likelihood ratio test comparing models with and without treatment-covariate interactions were compared. The test showed that the interactions did not add value meaning no significative heterogeneity was found in the IST dataset. The adequacy index was also computed to see how much predictive information was due to the treatment-covariate interactions. For IST, the adequacy index was equal to 0.993 meaning adding the interactions only accounted for 0.7\% of the predictive information. 
The ITRs had a PAPE close to $0$ meaning that the ITRs did not perform better than a rule that treated randomly the same proportion of patients. The PAPE values of most meta-learners were even slightly negative indicating that a non-individualized rule performed slightly better than those individualized rules. 
The values of $B_{pos}$  and $B_{neg}$ were close to 0, meaning there were not many benefits of giving the evaluated treatment to patients with a positive score or not giving the evaluated treatment to patients with a negative score.
The proportion of patients for whom aspirin was recommended by the different ITRs ranged from $0.114$ to $0.899$, with most methods producing an ITR that recommended the evaluated treatment for more than 50\% of patients. Methods belonging to the same family had similar proportions. Despite the significant disparity of proportions, the estimated values of the ITRS were similar showing that giving the evaluated treatment to more or fewer patients did not improve the value. For instance, OWL's ITR recommended treating $0.898$ of patients, and CWL's ITR recommended treating $0.114$ of patients but their rule's values were $0.399$ and $0.400$ respectively. The mean outcome when no one was treated ($0.380$) was close to the mean outcome when everyone was treated ($0.396$), which further implies that the treatment had a limited impact on the outcome on average. The mean outcome under the individualized rule was above the mean outcome when no one was treated for all methods. However, only five methods (Causal forests, A-learning, modified covariate method, OWL, and CWL) had mean outcome under the rule above the mean outcome when everyone was treated, and even for those methods, the mean outcome did not notably surpass the mean outcome if everyone is treated. Generally, the ITRs developed by the different methods did not drastically improve the mean outcome.\\

\begin{sidewaystable}
\footnotesize
\caption{Results of the metrics for each method applied to the IST dataset.\label{tab:res_ist}}
\centering
\begin{threeparttable}
\begin{tabular}{lllllllll}
\toprule
& $p_r$ & $\mathcal{V}(r)$ (SE) & $E(Y^0)$ (SE) & $E(Y^1)$ (SE) & $B_{pos}$ (SE) & $B_{neg}$ (SE) & PAPE (SE) & c for benefit (95\% CI)\\ 
\midrule
SL & 0.567 & 0.388 (0.009) & 0.380 (0.009) & 0.396 (0.009) & 0.014 (0.017) & -0.019 (0.018) & -0.002 (0.006) & 0.495 (0.476;0.514)\\
TL & 0.567 & 0.388 (0.009) & 0.380 (0.009) & 0.396 (0.009) & 0.014 (0.017) & -0.019 (0.019) & -0.002 (0.006) & 0.495 (0.476;0.514)\\
XL & 0.622 & 0.382 (0.009) & 0.380 (0.009) & 0.396 (0.009) & 0.004 (0.016) & -0.034 (0.020) & -0.008 (0.006) & 0.495 (0.477;0.513)\\
DRL & 0.622 & 0.382 (0.008) & 0.380 (0.009) & 0.396 (0.009) & 0.003 (0.016) & -0.035 (0.020) & -0.008 (0.006) & 0.495 (0.477;0.514)\\
RL & 0.626 & 0.388 (0.009) & 0.380 (0.009) & 0.396 (0.009) & 0.013 (0.016) & -0.022 (0.020) & -0.003 (0.006) & 0.501 (0.483;0.519)\\
SL RF & 0.526 & 0.395 (0.009) & 0.380 (0.009) & 0.396 (0.009) & 0.030 (0.017) & -0.004 (0.017) & 0.006 (0.006) & 0.498 (0.480;0.516)\\
TL RF & 0.515 & 0.390 (0.009) & 0.380 (0.009) & 0.396 (0.009) & 0.021 (0.018) & -0.013 (0.018) & 0.002 (0.006) & 0.498 (0.480;0.517)\\
XL RF & 0.543 & 0.393 (0.009) & 0.380 (0.009) & 0.396 (0.009) & 0.025 (0.017) & -0.008 (0.017) & 0.004 (0.006) & 0.504 (0.486;0.523)\\
DRL RF & 0.529 & 0.390 (0.009) & 0.380 (0.009) & 0.396 (0.009) & 0.020 (0.017) & -0.015 (0.018) & 0.001 (0.006) & 0.500 (0.482;0.518)\\
RL RF & 0.528 & 0.387 (0.009) & 0.380 (0.009) & 0.396 (0.009) & 0.018 (0.018) & -0.019 (0.018) & -0.001 (0.006) & 0.501 (0.482;0.519)\\
SL CF & 0.598 & 0.389 (0.009) & 0.380 (0.009) & 0.396 (0.009) & 0.018 (0.017) & -0.019 (0.019) & 0.000 (0.006) & 0.497 (0.478;0.515)\\
TL CF & 0.555 & 0.385 (0.009) & 0.380 (0.009) & 0.396 (0.009) & 0.009 (0.017) & -0.025 (0.019) & -0.004 (0.006) & 0.498 (0.480;0.516)\\
XL CF & 0.599 & 0.386 (0.009) & 0.380 (0.009) & 0.396 (0.009) & 0.011 (0.017) & -0.024 (0.018) & -0.004 (0.006) & 0.498 (0.480;0.516)\\
DRL CF & 0.552 & 0.386 (0.009) & 0.380 (0.009) & 0.396 (0.009) & 0.011 (0.016) & -0.023 (0.019) & -0.003 (0.006) & 0.499 (0.481;0.517)\\
RL CF & 0.548 & 0.389 (0.009) & 0.380 (0.009) & 0.396 (0.009) & 0.018 (0.018) & -0.016 (0.017) & 0.000 (0.006) & 0.501 (0.483;0.519)\\
PATH & 0.723 & 0.394 (0.009) & 0.380 (0.009) & 0.396 (0.009) & 0.020 (0.015) & -0.003 (0.014) & 0.002 (0.005) & 0.511 (0.493;0.529)\\
Causal Forests & 0.899 & 0.397 (0.008) & 0.380 (0.009) & 0.396 (0.009) & 0.019 (0.013) & 0.010 (0.032) & 0.002 (0.003) & 0.504 (0.486;0.522)\\
VT & 0.758 & 0.393 (0.009) & 0.380 (0.009) & 0.396 (0.009) & 0.017 (0.015) & -0.009 (0.019) & 0.000 (0.005) & ---\\
AL & 0.501 & 0.419 (0.008) & 0.380 (0.009) & 0.396 (0.009) & 0.089 (0.018) & 0.059 (0.018) & 0.031 (0.005) & 0.566 (0.548;0.583)\\
MCM & 0.501 & 0.419 (0.008) & 0.380 (0.009) & 0.396 (0.009) & 0.089 (0.017) & 0.059 (0.017) & 0.031 (0.005) & 0.566 (0.548;0.583)\\
OWL & 0.898  & 0.399 (0.009) & 0.380 (0.009) & 0.396 (0.009) & 0.022 (0.013) & 0.000 (0.017) & 0.004 (0.003) & ---\\
CWL & 0.114 & 0.400 (0.009) & 0.380 (0.009) & 0.396 (0.009) & 0.075 (0.034) & 0.007 (0.012) & 0.018 (0.004) & ---\\
\bottomrule
\end{tabular}
\begin{tablenotes}
\item $p_r$ refers to the proportion of patients for which the rule recommends treatment. $E(Y^0)$ and $E(Y^1)$ refer respectively to the mean outcome when no one is treated and the mean outcome when everyone is treated.SL: S-learner, TL: T-learner, XL: X-learner, DRL: DR-learner, RL: R-learner, RF: random forests, CF: cross-fitted, VT: virtual twins, MCM: modified covariate method, AL: A-learning, OWL: outcome weighted learning and CWL: contrast weighted learning.
\end{tablenotes}
\end{threeparttable}
\end{sidewaystable}

\begin{figure}
\centerline{\includegraphics[scale=1.7]{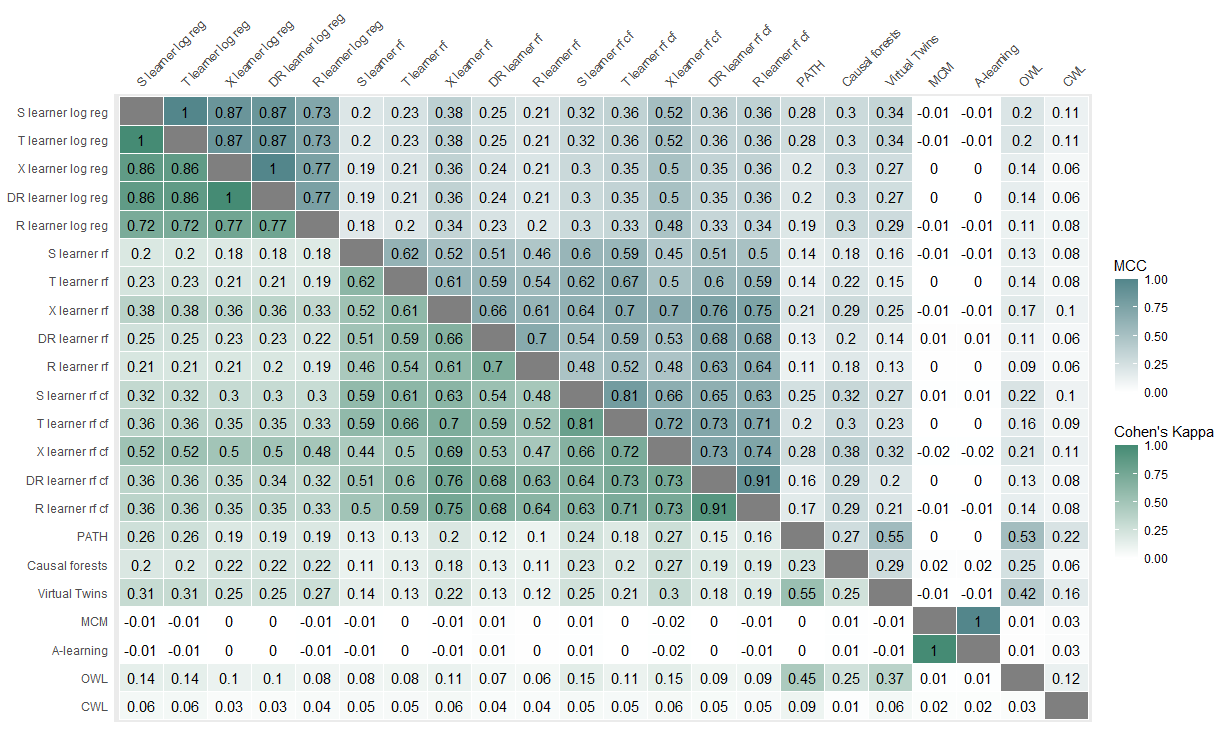}}
\caption{Heatmap representing the MCC and Cohen's Kappa for each combination of two ITRs using the International Stroke Trial.\label{fig:ist}}
\end{figure}

\noindent Overall, MCC and kappa's coefficient produced similar values (Figure~\ref{fig:ist}).
Most methods had considerable disagreements and thus almost no correlation regarding the people treated with the evaluated treatment in their rules which can indicate that the rules did not consider the same characteristics for the allocation of the treatment.
A better concordance was found among methods of the same family. For instance, the ITRs developed with the parametric meta-learners agreed to treat similar patients and had MCC and Cohen's kappa values ranging from $0.77$ to $1$. Similarly, non-parametric meta-learners had a positive moderate to high correlation with each other and with their crossfitted counterparts. However, they had less correlation with the parametric ones. 
A-learning and the modified covariate method generated the same ITR and therefore had coefficients of $1$.
The ITRs obtained with the different methods generally did not recommend the evaluated treatment to the same patients, which draws some concerns for their usability in practice.\\

\begin{figure}
\centerline{\includegraphics[scale=1.7]{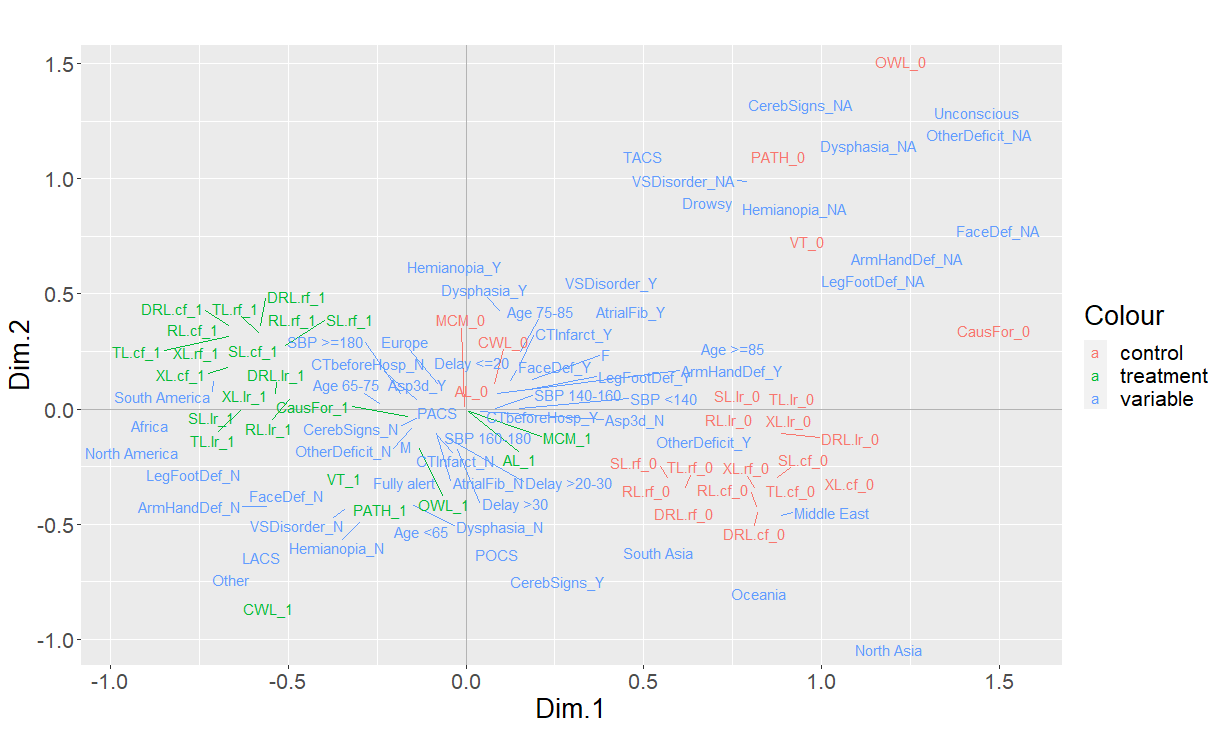}}
\caption{Multiple Correspondence Analysis on the International Stroke Trial showing all levels of each variable and the treatment recommendation of the individualized treatment rules.\label{fig:mca_ist}}
\end{figure}

\noindent A majority of characteristics were located near the origin and were not associated with the treatment allocation of the different ITRs (Figure~\ref{fig:mca_ist}). Virtual twins' ITR recommended not treating patients in a drowsy state and patients with a TACS stroke and recommended treating fully alert patients, younger patients, and patients with no deficit or disorder. Non-parametric meta-learners produced ITRs that recommended treatment for patients from South America and no treatment for patients from the Middle East, South Asia, and Oceania. CWL's ITR recommended treating patients with a LACS stroke or another type of stroke while OWL's ITR recommended not giving aspirin to unconscious patients.
The MCA was concordant with what was found in Figure~\ref{fig:ist} and reflected well the disagreement in terms of treatment allocation between the ITRs.

\subsection{CRASH-3}\label{ssec:crash3}
In a second stage, we compared the methods on the CRASH-3 dataset \cite{crash3_effects_2019}. Some heterogeneity in early treatment administration has been found in the CRASH-3 trial, therefore we thought it would be interesting to develop ITRs on this data \cite{crash3_effects_2019}. CRASH-3 is a multi-center randomized control trial consisting of $9,072$ patients from $175$ hospitals over $29$ countries. The aim of this trial was to examine the effects of tranexamic acid (TXA) in patients with acute traumatic brain injury. This paper used head injury death as the outcome (1=no and 0=yes). Six covariates were included in the methods: $2$ categorical variables and $4$ continuous variables. A binary treatment variable (0=Placebo and 1=TXA) was used. A description of the covariates and outcome is given in Supporting Material~S3.\\

\noindent Table~\ref{tab:res_crash3} shows the values of the metrics obtained with each method. The results of the training set are given in Supplementary Material~S3.
Recall that higher values of c-statistic for benefit are better but it is rare to obtain values above $0.6$ \cite{vanKlaveren2018}. Here, c-statistic for benefit values were all around or under $0.5$, indicating poor discrimination except for A-learning and the modified covariate method's ITRs which had higher values ($0.721$ and $0.720$ respectively). Excluding  A-learning and the modified covariate method's ITRs, the ITRs were not able to differentiate patients benefiting from taking the evaluated treatment from patients not benefiting.
The PAPE values were close to $0$ meaning that the ITRs did not perform better than a rule which randomly treated the same proportion of patients. There was a mix of positive and negative values but they all remained close to $0$.
A-learning and the modified covariate method's ITRs had $B_{pos}$ and $B_{neg}$ values around $0.2$, showing some benefits of giving the evaluated treatment to patients with a positive score and not giving the evaluated treatment to patients with a negative score, which was not the case for the other ITRs who had values near $0$. A-learning and the modified covariate method outperformed other approaches. This better performance is attributed to the treatment rules they developed, predominantly recommending treatment for patients with moderate Glasgow coma scores and reactive pupils. These patients’ profiles align with findings from the CRASH-3 study.
The proportions of people for which the treatment was recommended went from $0$ to $1$ with a majority of methods recommending to give the evaluated treatment to over 60\% of patients. Note that CWL's ITR chose to give the evaluated treatment to no one whereas OWL's ITR chose to give the evaluated treatment to everyone.
The rules' mean outcomes were almost identical and practically all above $0.8$, although the proportion of treated patients differed for each method leading us to conclude that there was a negligible treatment effect. This can be emphasized by looking at the mean outcome when no one is treated and the mean outcome when all the individuals are treated. The mean outcome when no one was treated ($0.802$) was close to the mean outcome when everyone was treated ($0.819$). Comparing these two mean outcomes to the mean outcome under the rules, we found that all the methods, except the crossfitted S-learner, had a mean outcome higher than the mean outcome when no one was treated, but only three of them (PATH, A-learning, and the modified covariate method) had a better mean outcome than the mean outcome when everyone was treated.\\

\begin{sidewaystable}
\footnotesize
\caption{Results of the metrics for each method applied to the CRASH-3 dataset.\label{tab:res_crash3}}
\centering
\begin{threeparttable}
\begin{tabular}{lllllllll}
\toprule
& $p_r$ & $\mathcal{V}(r)$ (SE) & $E(Y^0)$ (SE) & $E(Y^1)$ (SE) & $B_{pos}$ (SE)& $B_{neg}$ (SE)& PAPE (SE)& c for benefit (95\% CI)\\ 
\midrule
SL & 0.798 & 0.818 (0.009) & 0.802 (0.010) & 0.819 (0.010) & 0.019 (0.013) & 0.007 (0.040) & 0.002 (0.007) & 0.485 (0.448;0.522)\\
TL & 0.798 & 0.818 (0.009) & 0.802 (0.010) & 0.819 (0.010) & 0.019 (0.013) & 0.007 (0.037) & 0.002 (0.007) & 0.485 (0.448;0.522)\\
XL & 0.711 & 0.812 (0.010) & 0.802 (0.010) & 0.819 (0.010) & 0.013 (0.014) & -0.021 (0.031) & -0.002 (0.007) & 0.480 (0.447;0.514)\\
DRL & 0.711 & 0.812 (0.010) & 0.802 (0.010) & 0.819 (0.010) & 0.013 (0.015) & -0.021 (0.030) & -0.002 (0.007) & 0.481 (0.448;0.513)\\
RL & 0.711 & 0.812 (0.010) & 0.802 (0.010) & 0.819 (0.010) & 0.013 (0.014) & -0.021 (0.030) & -0.002 (0.007) & 0.481 (0.448;0.513)\\
SL RF & 0.496 & 0.804 (0.010) & 0.802 (0.010) & 0.819 (0.010) & 0.003 (0.019) & -0.030 (0.020) & -0.006 (0.007) & 0.489 (0.454;0.524)\\
TL RF & 0.514 & 0.805 (0.010) & 0.802 (0.010) & 0.819 (0.010) & 0.007 (0.018) & -0.022 (0.021) & -0.002 (0.007) & 0.493 (0.458;0.528)\\
XL RF & 0.579 & 0.809 (0.010) & 0.802 (0.010) & 0.819 (0.010) & 0.012 (0.017) & -0.022 (0.021) & -0.002 (0.007) & 0.490 (0.457;0.524)\\
DRL RF & 0.553 & 0.813 (0.009) & 0.802 (0.010) & 0.819 (0.010) & 0.019 (0.016) & -0.015 (0.022) & 0.001 (0.007) & 0.518 (0.484;0.552)\\
RL RF & 0.531 & 0.812 (0.010) & 0.802 (0.010) & 0.819 (0.010) & 0.018 (0.018) & -0.016 (0.021) & 0.001 (0.007) & 0.512 (0.479;0.545)\\
SL CF & 0.641 & 0.798 (0.010) & 0.802 (0.010) & 0.819 (0.010) & -0.008 (0.015) & -0.056 (0.027) & 0.001 (0.007) & 0.469 (0.435;0.504)\\
TL CF & 0.605 & 0.803 (0.010) & 0.802 (0.010) & 0.819 (0.010) & 0.000 (0.015) & -0.041 (0.025) & -0.009 (0.007) & 0.476 (0.443;0.510)\\
XL CF & 0.666 & 0.804 (0.010) & 0.802 (0.010) & 0.819 (0.010) & 0.003 (0.014) & -0.051 (0.028) & -0.010 (0.008) & 0.473 (0.440;0.507)\\
DRL CF & 0.593 & 0.803 (0.010) & 0.802 (0.010) & 0.819 (0.010) & 0.001 (0.017) & -0.039 (0.024) & -0.010 (0.007) & 0.485 (0.451;0.518)\\
RL CF & 0.592 & 0.802 (0.010) & 0.802 (0.010) & 0.819 (0.010) & -0.002 (0.017) & -0.042 (0.023) & -0.011 (0.007) & 0.486 (0.454;0.519)\\
PATH & 0.979 & 0.820 (0.009) & 0.802 (0.010) & 0.819 (0.010) & 0.019 (0.013) & -0.008 (0.023) & 0.001 (0.004) & 0.494 (0.462;0.526)\\
Causal Forests & 0.905 & 0.817 (0.010) & 0.802 (0.010) & 0.819 (0.010) & 0.016 (0.013) & 0.003 (0.058) & -0.001 (0.006) & 0.471 (0.437;0.506)\\
VT & 0.897 & 0.819 (0.010) & 0.802 (0.010) & 0.819 (0.010) & 0.019 (0.013) & 0.018 (0.050) & 0.002 (0.007) & ---\\
AL & 0.503 & 0.859 (0.007) & 0.802 (0.010) & 0.819 (0.010) & 0.236 (0.027) & 0.212 (0.027) & 0.049 (0.004) & 0.721 (0.699;0.743)\\
MCM & 0.503 & 0.859 (0.007) & 0.802 (0.010) & 0.819 (0.010) & 0.236 (0.027) & 0.212 (0.028) & 0.049 (0.004) & 0.720 (0.698;0.743)\\
OWL & 1  & 0.819 (0.010) & 0.802 (0.010) & 0.819 (0.010) & 0.017 (0.014) & --- & 0.000 (0) & ---\\
CWL & 0 & 0.802 (0.010) & 0.802 (0.010) & 0.819 (0.010) & --- & -0.017 (0.013) & 0.000 (0) & ---\\
\bottomrule
\end{tabular}
\begin{tablenotes}
\item $p_r$ refers to the proportion of patients for which the rule recommends treatment. $E(Y^0)$ and $E(Y^1)$ refer respectively to the mean outcome when no one is treated and the mean outcome when everyone is treated. SL: S-learner, TL: T-learner, XL: X-learner, DRL: DR-learner, RL: R-learner, RF: random forests, CF: cross-fitted, VT: virtual twins, MCM: modified covariate method, AL: A-learning, OWL: outcome weighted learning and CWL: contrast weighted learning.
\end{tablenotes}
\end{threeparttable}
\end{sidewaystable}

\begin{figure}
\centering
\includegraphics[scale=1.7]{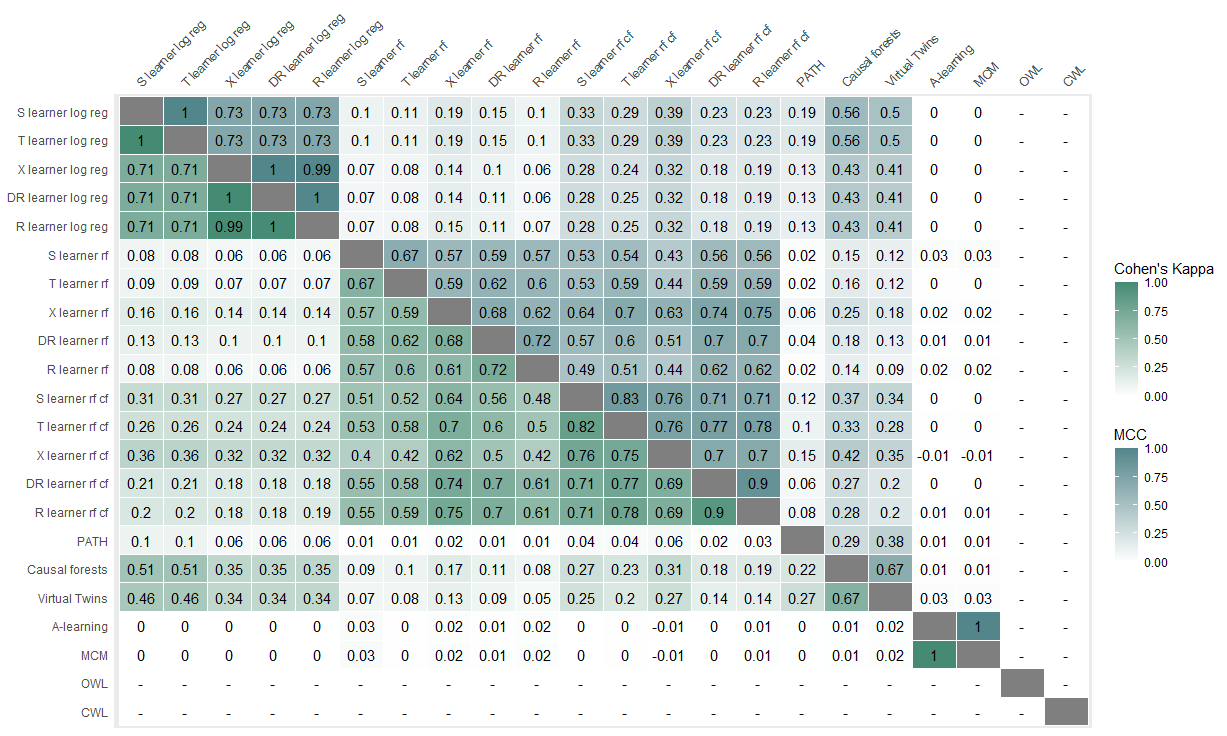}
\caption{Heatmap representing the MCC and Cohen's Kappa for each combination of two ITRs using the CRASH-3 trial.\label{fig:crash3}}
\end{figure}

\noindent MCC and Cohen's Kappa coefficient were concordant and gave coefficients of similar magnitude (Figure~\ref{fig:crash3}). When one of the ITRs recommended treating everyone or no one with the evaluated treatment, it did not make sense to calculate the MCC and Cohen's Kappa coefficient, therefore we put a dashed line in those cases.
Parametric meta-learners had a strong concordance with each other with high coefficients. The same thing was observed for non-parametric meta-learners whether crossfit was applied or not, as well as for A-learning and the Modified covariate method. As for the IST, a strong concordance is only found between methods belonging to the same family (e.g. parametric meta-learners, non-parametric meta-learners, A-learner and the modified covariate method). Otherwise, the correlation between the ITRs was moderate and most of the time low. The ITRs did not recommend the evaluated treatment to the same patients. The choice of the method had a big impact on the treatment allocation, meaning that in practice two different methods could lead to completely different rules. \\

\begin{figure}
\centerline{\includegraphics[scale=1.7]{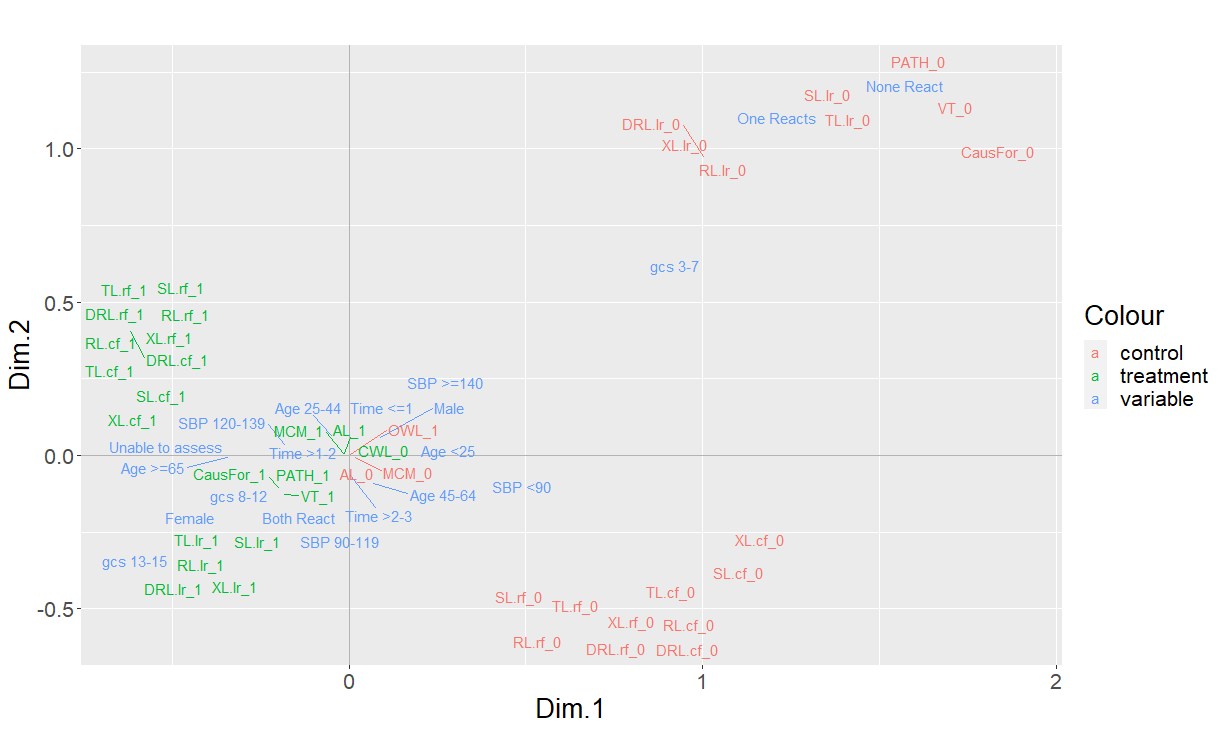}}
\caption{Multiple Correspondence Analysis of the CRASH-3 trial representing the variables' levels and the treatment recommendation of every individualized treatment rule.\label{fig:mca_crash3}}
\end{figure}

\noindent The ITRs developed with the parametric meta-learners, virtual twins and causal forests recommended not to treat patients with a low Glasgow Coma Scale score or/and patients with none or only pupil that reacted, whereas they recommended the treatment to patients with a moderate to high Glasgow Coma Scale score, patients who were female and patients with moderate systolic blood pressure (Figure~\ref{fig:mca_crash3}). The non-parametric meta-learners' ITRs recommended treating patients younger patients with relatively high blood pressure and not treating patients with low blood pressure. The MCA reflected well the agreement results that were found in Figure~\ref{fig:crash3}. Akin methods' ITRs agreed on the treatment allocation but overall the ITRs did not take into account the same characteristics for the treatment decision.

\section{Discussion}\label{sec:discuss}

This paper compared different methods used to construct individualized treatment rules using data from two RCTs: the International Stroke Trial and the CRASH-3 trial. We considered 22 methods belonging to two different classes. The first class included methods that predicted the ITE to derive an ITR: meta-learners (T-learner, S-learner, X-learner, DR-learner and R-learner, both with logistic regression or random forests as a base learner with and without cross-fit), PATH, causal forests, virtual twins, A-learning and modified covariate method. The second class covered methods that directly estimated the ITR without explicitly estimating ITEs: outcome-weighted learning and contrast weighted learning. For each trial, the performance of the ITRs was assessed with various metrics. The pairwise agreement between ITRs was also evaluated. 

\noindent Results showed that the ITRs obtained by the different methods generally had considerable disagreements regarding the individuals to be treated with the evaluated treatment for both trials. The proportions of patients for whom the evaluated treatment was recommended by the rules were very different depending on which method was employed to build the ITR and the  Cohen’s kappa and Matthews correlation coefficients were low. A better concordance was found among methods of the same family (e.g. among all meta-learners with parametric models, or all meta-learners with non-parametric models and cross-fitting). Overall, when evaluating the performance of ITRs in a hold-out validation sample (33\% of the original sample selected at random), results showed that all ITRs had limited performance, whatever the performance in the training set, which suggests a high potential of optimism for the algorithms.

\noindent The limited performance results might be due to the distribution of treatment effects and the level of heterogeneity. Although some heterogeneity of treatment effects was found in the trials used in this work, especially in the CRASH-3 trial \cite{crash3_effects_2019}, the level of heterogeneity might not be sufficient to develop a beneficial individualized treatment rule. This result was reinforced by performing likelihood ratio tests and calculating adequacy indexes. For both trials, the likelihood ratio tests led to the conclusion that there was no evidence of significant heterogeneity and the values of the adequacy indexes showed that the treatment-covariate interactions only accounted for a low percentage of the predictive information. 
Another explanation for the limited performance might be the sample size. Even if the methods were compared on two large RCTs, perhaps more data is needed to obtain a better performance. A solution might be using individual participant data meta-analyses (IPD-MA) since they include a larger number of participants. However, one should consider the heterogeneity that may arise between the studies included in the meta-analysis. Different methods to tackle the heterogeneity in IPD-MA have been proposed and compared in previous works \cite{brionbouvier:hal-03735613,Steyerberg2019,Debray2013}. 

\noindent In a previous work, Rekkas et al. \cite{rekkas_estimating_2023} demonstrated via a simulation study that "complex" methods, which are more flexible, require large sample sizes to perform well and that, when one has access to moderate sample sizes, simpler risk modeling methods recommended by the PATH statement \cite{Kent_path} should be preferred to obtain a good performance. Using more parsimonious models with fewer covariates, like what has been done for the SYNTAX II score, might also lead to more robust ITRs with better agreements \cite{takahashi_redevelopment_2020}. 
Investigating for which distribution of treatment effects, a model can have good discrimination, and thus be able to develop a beneficial ITR, as well as the requirements in effective sample size to allow reliable development of ITRs, is worth studying.

\noindent Some comparisons of methods used to construct ITRs have been conducted in the past \cite{jacob_cate_2021,zhang_unified_2021}, but to our knowledge, no study has investigated the agreements in terms of the treatment decision with all the methods presented in this project. Both Jacob and Zhang et al. have found that the methods had different performances \cite{jacob_cate_2021,zhang_unified_2021}. These results were concordant with ours. 

\noindent Although we compared many methods in this work, we did not include every existing method. Indeed, we decided to focus on methods that are commonly used and that are easily computed or for which an R package was available. We also focused on real data, and a simulation study should be conducted to better delineate the parameters associated with a better performance of the methods. A recent simulation study showed that the sample size and the shape of the distribution of treatments impacted the performance of the methods, particularly the performance of "complex" methods \cite{rekkas_estimating_2023}. However, we considered that the illustration on two large RCTs was necessary to study the agreement between the different ITRs in real settings, simulations being often over-simplified. Using real data also allows for tailoring each method, in the sense that each model does not necessarily need to have the same variables.  
Furthermore, in this paper, we decided to compare the ITRs' decisions on randomized controlled trials. Constructing ITRs can also be done with observational data. Observational databases have the potential to include much more participants, and more diverse participants, than trials, and thus might have both more heterogeneity and larger sample sizes, and could be a better source of data to develop ITRs in practice. Although an effort was dedicated to method optimization, specifically optimizing hyperparameters for tree-based methods through cross-validation to maximize accuracy, it is plausible that further optimization might have led to improved method performance.

\noindent In conclusion, the significant disagreements that the methods had regarding the treatment allocation suggest that the methods are not interchangeable. Therefore, the chosen method greatly influences the patients for which the evaluated treatment is recommended. It draws some concerns about their practical use. Some ITRs have been developed in the past using one method with similar RCTs \cite{Nguyen2020}. Using multiple methods and comparing the obtained ITRs, as suggested by Jacob, might be a solution when one wants to develop an ITR in practice \cite{jacob_cate_2021}. However, in most cases, more simple approaches such as the risk modeling method advocated in PATH \cite{Kent_path}, or carefully adding specific interactions between prespecified treatment-effect modifiers and treatment in the model, as done in the revised SYNTAX score II \cite{takahashi_redevelopment_2020} may be a better strategy than currently available ITR  algorithms which may be misleading by overfitting the heterogeneity of treatment effects. Also, methods that allow evaluating the model calibration for benefit may be favored. Evaluating a priori the probability of identifying a beneficial ITR, as suggested by Cain et al. \cite{cain_design_2022}, might also be taken under consideration.

\section*{Declarations}
\subsection*{Competing interests}
The authors declare that they have no competing interests.
\subsection*{Funding}
FB and RP acknowledge support by the French Agence Nationale de la Recherche as part of the “Investissements d’avenir” program, reference ANR-19-P3IA-0001 (PRAIRIE 3IA Institute). This work was partially funded by the Agence Nationale de la Recherche, under grant agreement no. ANR-18-CE36-0010-01.
\subsection*{Authors' contributions}
Study concept and design: FB, CS, FP and RP. Analysis and interpretation of data: All authors. Drafting of the manuscript: FB, FP, and RP. Critical revision of the manuscript for important intellectual content: All authors.
\subsection*{Acknowledgements}
The authors wish to thank the investigators of the International Stroke Trial and the CRASH-3 trial for providing the datasets compared in this work. 
The authors would also like to thank Viet-Thi Tran and François Grolleau for contributing to the comparison using the CRASH-3 dataset. 
Finally, We thank the two anonymous referees for their comments that greatly improved the paper.
\subsection*{Data availability statement}
The data that support the findings of this study are openly available in IST dataset at \\
https://datashare.ed.ac.uk/handle/10283/124 \\
and in CRASH-3 at https://freebird.lshtm.ac.uk/index.php/available-trials/

\clearpage
\bibliographystyle{ama}
\bibliography{ITR}

\end{document}